# Pegs, Floats, and Forests: A Machine Learning Revisit of Exchange Rate Regimes and Growth in Transition Economies

**Marjan Petreski**
University American College Skopje, North Macedonia
Finance Think – Economic Research  Policy Institute, Skopje, North Macedonia
PEP – Partnership for Economic Policy, Nairobi, Kenya
marjan.petreski@uacs.edu.mk

**Abstract**

This paper combines traditional panel econometrics with random forest machine learning to revisit the relationship between exchange rate regimes and economic growth for 27 transition economies over 1991–2019. Exploiting the Couharde-Grekou (2024) probabilistic synthesis classification, the random forest approach non-parametrically confirms and sharpens what fixed-effects and system GMM estimation establish parametrically: intermediate exchange rate regimes consistently underperform fixed arrangements, with growth penalties ranging from −1.0 to −10.4 percentage points, while floating regimes show negative but largely insignificant differentials. Beyond regime effects, the machine learning analysis reveals that the intermediate regime penalty is sharpest precisely where institutions are weakest — non-parametric validation that institutional capacity, not regime label alone, determines whether exchange rate anchoring pays off. The regime-growth relationship is further concentrated in the pre-2003 stabilization era and is absent among EU member economies, suggesting the growth dividend from exchange rate anchoring eroded as institutional convergence advanced. Together, these findings demonstrate how machine learning variable importance metrics can corroborate and enrich causal inference from panel methods, while supporting the view that exchange rate anchoring carried a meaningful credibility dividend during the formative phase of transition.

## 1. Introduction

The relationship between exchange rate regimes (ERR) and economic growth has occupied a prominent place in the international macroeconomics literature for decades. Yet, despite extensive theoretical and empirical investigation, no consensus has emerged. Some studies find that pegged regimes support growth by reducing policy uncertainty and fostering trade and investment; others conclude that flexible regimes outperform pegs by providing an adjustment mechanism against external shocks; and a third group finds no systematic relationship at all (Petreski, 2009). This inconclusiveness reflects genuine theoretical ambiguity, but also a set of persistent empirical challenges: inappropriate growth frameworks, inadequate treatment of endogeneity, sample-selection bias, and — critically — the difficulty of measuring exchange rate regimes themselves.

Transition economies of Central and Eastern Europe (CEE), Southeastern Europe (SEE) and the Commonwealth of Independent States (CIS) represent a particularly compelling laboratory for studying this relationship. When these economies began their departure from central planning in the early 1990s, they inherited a set of structural conditions that made the exchange rate regime both more consequential and harder to identify than in other country groups. Repressed inflation, large fiscal deficits, currency inconvertibility, underdeveloped financial systems, and weak trade linkages with market economies all characterized the starting point (Sachs, 1996). As price liberalization unleashed inflationary pressure, many countries sought exchange rate anchors to impose monetary credibility — the Visegrad group and the Baltic states being prominent examples — while others adopted floats alongside stringent stabilization programs. Meanwhile, capital account liberalization, banking sector reform, interest rate deregulation, and rapid trade opening fundamentally altered the macroeconomic environment throughout the 1990s and 2000s, creating conditions under which the de facto exchange rate regime pursued in practice often diverged from what was officially declared (Calvo and Reinhart, 2002).

This context has direct implications for how the ERR-growth nexus should be studied in transition economies. Petreski (2014) demonstrated that three prominent de facto classification systems — those of Levy-Yeyati and Sturzenegger (2005), Reinhart and Rogoff (2004), and Bubula and Otker-Robe (2002) — disagreed substantially in their classification of transition economies, and that this disagreement was itself a product of transition-specific characteristics: high inflation episodes drove classification disagreement in the early transition period, while trade openness and interest rate liberalization drove it in the later period. Once these conflicting observations were removed and selectivity bias corrected for, all three systems converged on the same qualitative finding — that pegs and intermediate regimes significantly outperformed floats in terms of economic growth. The implication was clear: measurement matters enormously, and the transition context complicates measurement in ways that standard approaches do not adequately handle.

Two developments motivate a revisitation of these findings. First, the classification landscape has advanced significantly. Couharde and Grekou (2024) propose a synthesis classification that combines information from multiple existing de facto classification systems into a probabilistic framework, assigning to each country-year observation not a hard categorical label but a vector of probabilities of belonging to the fixed, intermediate, and floating categories. This approach is particularly well-suited to the transition context, where regime behavior was genuinely ambiguous and where the disagreement among classifiers documented in Petreski (2014) was itself informative. Rather than resolving classification uncertainty by discarding conflicting observations, the synthesis classification retains and quantifies that uncertainty, allowing it to enter the empirical analysis directly. Second, the time horizon has expanded materially. Earlier papers rarely cover the Global Financial Crisis, the subsequent sovereign debt turbulence, and the phase of EU integration for CEE economies. The Couharde and Grekou (2024) data extend to 2019 and provide for capturing of these developments. This longer window captures fundamentally different macroeconomic conditions and allows assessment of whether the ERR-growth relationship has evolved as transition economies matured and diverged.

Beyond the classification advance, this paper introduces a methodological novelty by applying machine learning techniques — specifically, random forest estimation — to identify the country characteristics that condition the ERR-growth relationship in transition economies. Parametric frameworks usually employed in earlier studies, while carefully specified, necessarily imposes linearity and requires pre-specified interactions between regime type and conditioning variables. Random forests relax these constraints, allowing the data to reveal non-linearities and interactions without prior theoretical commitment. Variable importance measures from the random forest estimation complement the regression-based results by providing a data-driven ranking of the factors — including transition-specific ones such as inflation history, credit development, trade openness, and financial liberalization — that most powerfully mediate the growth effects of different exchange rate arrangements.

The remainder of the paper is organized as follows. Section 2 briefly reviews the theoretical and empirical literature. Section 3 describes the underlying methodology and the Couharde and Grekou (2024) classification, as well provides a descriptive analysis of exchange rate regime patterns in transition economies over 1991–2019. Section 4 presents the growth regression framework and results using both the categorical and probabilistic dimensions of the synthesis classification. Section 5 reports the random forest analysis and variable importance findings. Section 6 concludes.

## 2. Exchange Rate Regimes and Growth: State of the Literature

### 2.1 Theoretical Channels

Economic theory does not offer an unambiguous prediction about how the exchange rate regime affects long-run economic growth. As a nominal variable, the exchange rate might be expected to leave real outcomes unchanged in the long run — a view consistent with the natural rate hypothesis and the classical dichotomy (Goldstein, 2002). Yet a substantial body of theoretical work argues that the regime choice exerts real effects through at least three channels: trade, investment, and productivity.

The trade channel operates through the uncertainty imposed by exchange rate flexibility. Advocates of pegged regimes argue that by eliminating bilateral exchange rate risk, a peg promotes trade volumes and predictability, which in turn supports growth through specialization, technology transfer, and scale economies (Gylfason, 2000; Moreno, 2001). De Grauwe and Schnabl (2004) extend this argument, emphasizing that a credible peg lowers the country risk premium embedded in interest rates, simultaneously stimulating consumption and investment. Conversely, opponents note that flexible rates need not depress trade if agents can hedge through forward markets, though for developing economies where such instruments are absent, the relationship between exchange rate volatility and trade remains empirically contested (Viaene and de Vries, 1992; Nilsson and Nilsson, 2000).

The investment channel relates regime choice to uncertainty more broadly. Bohm and Funke (2001) argue that reduced uncertainty under a peg promotes investment by removing the "wait and see" attitude that exchange rate volatility induces in forward-looking firms. Dixit (1989) formalizes this through hysteresis: instability leads to disinvestment or postponement of planned investment. However, the regime-investment link remains empirically thin, partly because investment decisions depend on many factors beyond the exchange rate environment and partly because the direction of causality is difficult to establish cleanly.

The productivity channel is perhaps the most theoretically nuanced. Ghosh et al. (1997) argue that while pegs may enhance investment, floats produce faster productivity growth, leaving the net effect on output ambiguous. Aghion et al. (2009) offer a more precise mechanism: in economies with underdeveloped credit markets, a negative aggregate shock under a peg depresses firm profitability to the point where long-term productivity-enhancing investments cannot be financed. The implication is that the growth effects of exchange rate flexibility are conditioned by financial sector depth — countries with shallow financial systems may benefit more from pegging, while those with developed credit markets gain from flexibility. Slavtcheva (2015) provides direct empirical support for this interaction, finding that countries with limited financial development experience positive growth effects from fixed regimes. This financial development conditionality offers a compelling theoretical explanation for the divergent empirical findings across income and development groups, and has direct relevance for transition economies where financial systems were rebuilt from scratch during the 1990s.

A flexible exchange rate also serves as an absorber of external shocks. Friedman (1953) established the classical case: when prices are sticky, a flexible rate allows faster real adjustment than deflation or wage cuts under a peg. Bailliu et al. (2003) translate this into growth terms, arguing that smoother adjustment to shocks keeps the economy operating closer to capacity on average. Fisher (2001) adds that in a world of high capital mobility, defending a peg under speculative attack imposes severe interest rate and reserve costs, making pegs increasingly fragile and potentially growth-destructive in crisis episodes — a concern particularly relevant for the post-2008 period.

Reconciling these competing channels, the theoretical literature reaches no clean verdict. The sign of the regime-growth relationship depends on which channel dominates, and this in turn depends on structural features — financial development, trade openness, shock exposure, institutional quality, and labour market flexibility — that vary substantially across countries and over time. This conditional nature of the relationship has progressively moved to the center of the empirical debate, as we now discuss.

### 2.2 Empirical Evidence: Persistent Divergence

The empirical literature on exchange rate regimes and growth is extensive and its findings remain strikingly divergent despite decades of increasingly sophisticated analysis. Early descriptive work by Baxter and Stockman (1989) found little systematic relationship between exchange rate arrangements and real macroeconomic variables, while Mundell (1995) found considerably higher growth under the generalized pegging of the Bretton Woods era. The absence of econometric identification in both studies limited their interpretability, but they established the basic empirical puzzle that subsequent work has attempted to resolve.

The first generation of regression-based studies produced equally mixed results. Ghosh et al. (1997) found slightly higher GDP growth under floating regimes but the highest growth under intermediate arrangements — a pattern that has recurred throughout the literature. Moreno (2000, 2001) found real growth higher under pegs by 1.1 and 3 percentage points respectively in developing and East Asian economies, though the difference narrowed once survivor bias was addressed.

The major econometric contributions of the 2000s sharpened the debate without resolving it. Levy-Yeyati and Sturzenegger (2003), using their own de facto classification for 183 countries, found that pegging was associated with slower growth in developing economies but had no effect in advanced ones. Husain et al. (2005), using the Reinhart-Rogoff classification, found that neither pegs nor floats systematically dominated, with regime durability mattering more than regime type per se. Rogoff et al. (2003) similarly found that growth implications depended heavily on the level of development, with more advanced economies benefiting from flexibility. Bleaney and Francisco (2007) found slower growth under rigid regimes in developing economies, while Dubas et al. (2005) found that de facto fixers grew approximately one percentage point faster than de facto floaters among non-industrial economies. These studies were reviewed comprehensively in Petreski (2009), who identified four structural flaws running through the empirical literature: inadequate growth frameworks, endogeneity of regime choice, sample selection bias, and the peso problem.

Post-2009 empirical work has not resolved the divergence but has refined the questions being asked. Among studies confirming significant regime effects for developing economies, Cruz-Rodriguez (2022), examining 125 countries over the post-Bretton Woods period, finds that developing countries with fixed regimes tend to experience higher growth — a conclusion corroborated by Ashour and Chen Yong (2018), who document a growth premium for pegged arrangements across 32 developing countries over 1974–2006 using standard panel estimators, and by Mohammed et al. (2017), whose cross-country analysis similarly finds fixed regimes associated with more stable and higher growth outcomes in economies lacking independent monetary credibility. Baycan (2016) addresses the endogeneity of regime choice directly through propensity score matching, pairing countries with similar observable characteristics but different exchange rate arrangements; the finding that fixed regimes continue to outperform floats after controlling for observable selection strengthens the causal interpretation of average-effect estimates. Bleaney, Saxena, and Yin (2018) redirect attention from average growth differentials toward tail outcomes, documenting that the growth consequences of regime choice are concentrated in collapse episodes: pegged regimes experience larger but less frequent growth collapses when devaluations occur, so that the average performance advantage of fixed arrangements coexists with episodically severe vulnerability.

A more prominent strand of recent research has shifted attention from average regime effects toward heterogeneity and conditionality. Slavtcheva (2015) demonstrates that the growth-enhancing effect of fixed regimes is concentrated in countries with limited financial development, consistent with the Aghion et al. (2009) mechanism. Kuokštis et al. (2025) introduce labour market flexibility as a further moderating variable, finding for a panel of 194 countries over 1970–2019 that fixed regimes hinder growth when labour markets are highly rigid but boost growth when they are flexible — an interaction that standard regression specifications with regime dummies alone cannot detect. The persistent inconclusiveness of average-effect estimates has thus given way to a recognition that the ERR-growth relationship is fundamentally heterogeneous, mediated by structural and institutional characteristics that interact with regime type in complex, potentially non-linear ways.

On the classification front, Ilzetzki, Reinhart, and Rogoff (2019) updated the original Reinhart-Rogoff algorithm with an extended dataset covering exchange arrangements through the early 21st century, documenting that the prevalence of de facto fixed regimes is considerably higher than official declarations suggest and that the US dollar has become ever more central as an anchor currency globally. Their subsequent review (Ilzetzki, Reinhart, and Rogoff, 2022) surveys the evolution of the global exchange rate system and the methodological challenges in classifying regimes, noting that disagreement among classification systems remains a fundamental problem. Eichengreen and Razo-Garcia (2013) had earlier documented systematically that the three major de facto classifications diverge in their assessments, and that these divergences likely drive at least part of the conflicting growth findings. This recognition of classification uncertainty as a substantive empirical problem, rather than a mere data nuisance, motivates the probabilistic approach of Couharde and Grekou (2024).

### 2.3 Evidence from Transition Economies

Transition economies occupy a distinct position in this literature, both as a theoretically interesting subgroup and as a context in which the standard empirical challenges are especially acute. When these economies departed from central planning in the early 1990s, they inherited structural conditions that made the exchange rate regime simultaneously more consequential and harder to classify: repressed inflation, large fiscal imbalances, currency inconvertibility, underdeveloped financial systems, and weak trade and financial linkages with market economies (Sachs, 1996). As price liberalization unleashed inflationary pressure, many countries sought exchange rate anchors to impose monetary credibility — the Visegrad group and the Baltic states being prominent examples — while others, including Bulgaria, Romania, Russia and Ukraine, adopted floats alongside stabilization programs. The tension between rapid structural change and inherited rigidity created persistent pressures on exchange rate behavior, causing many transition economies to exhibit actual exchange rate policies at variance with their official declarations (Dean, 2003; Calvo and Reinhart, 2002).

Domac et al. (2001) were among the first to examine the regime-growth nexus specifically for transition economies, using a switching regression approach that modeled each regime separately to address the Lucas critique and sample selection simultaneously. Their findings were inconclusive — an association between regime and growth was detected, but no regime emerged as systematically superior. De Grauwe and Schnabl (2004), examining ten CEE countries over 1994–2002 with GMM estimation, found that exchange rate pegging promoted growth, with results stronger than those from broader country samples. They attributed this to the particular importance of nominal stability in economies still building monetary credibility and converging toward European integration. Khan, Kebewar, and Nenovsky (2013) compared the performance of currency board arrangements and inflation targeting regimes in Eastern Europe from 2000 onwards, finding that the two monetary frameworks generated markedly different inflation-output dynamics, with currency board countries exhibiting lower inflation uncertainty but less capacity to absorb output shocks — a trade-off that became acutely visible during the Global Financial Crisis.

The GFC represented a critical test of exchange rate arrangements in transition economies. The IMF's European Department (Belhocine et al., 2016) documented that across Central, Eastern, and Southeastern Europe, the boom-bust cycle of 2003–2012 was generally less pronounced in economies with flexible exchange rates, which experienced more muted pre-crisis expansion and faster post-crisis recovery. However, this aggregate finding concealed important heterogeneity: fixed regimes, particularly currency boards, offered credibility and financial stability benefits that proved valuable in economies where euroization and inflation memories made floating politically and economically costly. The GFC thus reinforced the view that no single regime dominates across all circumstances, and that the transition economies themselves had diverged into structurally distinct groups — EU-integrated economies converging toward the euro, CIS economies pursuing varied and sometimes volatile monetary strategies, and Western Balkan economies in intermediate stages of institutional development.

Petreski (2014) addressed the classification problem directly, examining whether disagreements among the three major de facto classification systems were themselves products of transition-specific characteristics. Using panel probit estimation for 28 transition economies over 1991–2007, the paper found that high inflation caused classification disagreement in the early transition period, while trade openness and interest rate liberalization drove disagreement later. After removing conflicting observations and correcting for selectivity bias using the Heckman two-step procedure, all three systems converged on the finding that pegs and intermediate regimes significantly outperformed floats, with growth differentials of 2 to 4 percentage points on average. This result established both a substantive finding — fixed and intermediate arrangements support growth in transition economies — and a methodological insight: that classification disagreement in this context is not random noise but a systematic product of structural transition features.

While the contributions reviewed above — particularly Baycan (2016) on selection bias and Bleaney et al. (2018) on growth-collapse episodes — advance causal identification and outcome heterogeneity respectively, three dimensions remain unaddressed in combination and motivate the present paper. First, the post-2007 period is entirely absent from the earlier transition economy evidence, yet it includes the most consequential macroeconomic events since transition began: the Global Financial Crisis, the sovereign debt turbulence in the European periphery, the divergence between EU-integrated and the rest of transition economies, and several episodes of sharp exchange rate realignment including the Russian rouble crisis of 2014–15 and the associated pressures on neighboring economies. Whether the finding that fixed and intermediate regimes outperform floats survives in this more turbulent environment, and whether it holds equally across the now-divergent subgroups of transition economies, remains an open question. Second, all four recent studies treat regime assignment as a hard binary or trinomial classification, discarding the uncertainty that is especially prevalent in transition settings; the earlier transition economy work addressed this by removing conflicting observations, which is subtractive rather than informative. The synthesis classification of Couharde and Grekou (2024), which assigns regime probabilities rather than hard categorical labels, offers a fundamentally different approach: retaining all observations and allowing regime ambiguity to enter the analysis directly as information. Third, none of these studies applies machine learning methods to characterise regime-conditioned growth environments without pre-specifying interactions — a methodological gap that the random forest analysis of this paper fills. These three dimensions — temporal coverage, probabilistic classification, and non-parametric heterogeneity analysis — define the contribution of the present paper and directly motivate the design choices described in Section 3.

## 3. Methodology and Data

### 3.1 Empirical Framework

The empirical strategy proceeds in three steps of increasing methodological ambition. First, a standard growth regression with categorical exchange rate regime dummies is estimated as a conventional benchmark, allowing comparison with the existing literature. Second, the categorical dummies are replaced with the continuous regime probabilities provided by the Couharde and Grekou (2024) synthesis classification, capturing regime uncertainty directly as a continuous regressor. Third, and centrally, random forest estimation is applied to identify which country characteristics most powerfully condition the ERR-growth relationship, without imposing functional form or pre-specifying interactions. The first step anchors the analysis in familiar territory; the second exploits the distinctive feature of the CC classification; the third constitutes the primary methodological contribution.

### 3.2 Benchmark Growth Regression

The benchmark specification follows the standard growth regression framework of Barro and Sala-i-Martin (2004):

$$gdppc_{it} = \alpha + \beta_1 gdp90_i + \beta_2 gc_{it} + \beta_3 inv_{it} + \beta_4 inf_{it} + \beta_5 ttg_{it} + \beta_6 popgr_{it} + \beta_7 lpop_{it} + \beta_8 trade_{it} + \gamma_1 floating_{it} + \gamma_2 intermediate_{it} + u_i + \varepsilon_{it} \quad (1)$$

where $gdppc_{it}$ is GDP per capita growth; $gdp90_i$ captures conditional convergence through initial income; $gc_{it}$, $inv_{it}$, $inf_{it}$, $ttg_{it}$, $popgr_{it}$, $lpop_{it}$, and $trade_{it}$ are government consumption, investment, inflation, terms of trade growth, population growth, population size, and trade openness respectively; $floating_{it}$ and $intermediate_{it}$ are dummy variables derived from the SC3w three-way synthesis classification of Couharde and Grekou (2024), with fixed as the reference category; $u_i$ is a country fixed effect; and $\varepsilon_{it}$ is the idiosyncratic error.

The equation is estimated by panel fixed effects. Instrumental variable estimation using lagged values of potentially endogenous regressors is presented alongside, with instrument validity assessed via the Hansen test. Standard errors are clustered at the country level throughout.

### 3.3 Probabilistic Regime Specification

The synthesis classification uniquely provides, for each country-year observation, the posterior probabilities $Prob_Fixed_{it}$, $Prob_Intermediate_{it}$, and $Prob_Floating_{it}$, reflecting the degree of consensus among the twelve constituent classification systems. These probabilities sum to unity, so the probabilistic specification replaces the categorical dummies with $Prob_Floating_{it}$ and $Prob_Intermediate_{it}$, with $Prob_Fixed_{it}$ as the implicit reference:

$$gdppc_{it} = \alpha + \beta_1 gdp90_i + \beta_2 gc_{it} + \beta_3 inv_{it} + \beta_4 inf_{it} + \beta_5 ttg_{it} + \beta_6 popgr_{it} + \beta_7 lpop_{it} + \beta_8 trade_{it} + \gamma_1 Prob_Floating_{it} + \gamma_2 Prob_Intermediate_{it} + u_i + \varepsilon_{it} \quad (2)$$

The coefficients $\gamma_1$ and $\gamma_2$ capture the marginal growth effect of increasing certainty of a floating or intermediate regime respectively. A country-year with $Prob_Fixed = 0.9$ and one with $Prob_Fixed = 0.5$ are treated as genuinely different observations, reflecting the intuition that regime credibility and commitment — not merely regime type — matter for growth outcomes. This is particularly relevant for transition economies, where exchange rate behavior was frequently ambiguous and where the synthesis probabilities are often well away from the corners of the simplex. Estimation proceeds identically to the benchmark specification.

### 3.4 Random Forest Estimation

#### 3.4.1 Approach

The regression specifications in Sections 3.2 and 3.3 deliver average partial effects under parametric assumptions. They cannot detect interactions between regime type and conditioning variables unless those interactions are pre-specified, and they impose linearity throughout. Random forests (Breiman, 2001) relax both constraints. As an ensemble method that aggregates predictions across a large number of decision trees — each grown on a bootstrap sample of the data using a random subset of predictors at each split — random forests capture non-linearities and interactions automatically, without any prior theoretical commitment.

The key output for our purposes is the variable importance measure: the mean decrease in prediction accuracy when each variable is randomly permuted across all trees. Variables that consistently drive splits high in the tree hierarchy — reducing prediction error most — receive high importance scores. This importance ranking is entirely data-driven and reflects the joint predictive contribution of each variable, including through interactions with other predictors.

We use random forests to address the following question: among the candidate conditioning variables, which most powerfully predict growth outcomes, and does this ranking differ across exchange rate regime types? Estimating separate forests for fixed, intermediate, and floating regime subsamples and comparing the resulting importance rankings delivers a data-driven characterization of how the growth-conditioning environment differs across regimes. We additionally estimate a pooled forest that includes the ERR regime dummies as predictors alongside the other conditioning variables, directly assessing how regime type ranks relative to structural country characteristics in predicting growth.

#### 3.4.2 Predictor set

The predictor set for the random forest includes all variables from the benchmark regression (initial income, government consumption, investment, inflation, terms of trade growth, population growth, population size, and trade openness), augmented in three directions. First, and centrally, the three regime probability variables from the Couharde-Grekou synthesis classification — P(Fixed), P(Intermediate), and P(Floating) — are included as continuous predictors. This exploits the full probabilistic information on regime assignment rather than collapsing it to a hard categorical dummy, and constitutes the methodological bridge between the probabilistic regression specification of Section 3.3 and the non-parametric forest analysis: the forest can detect non-linear and interactive effects of regime certainty and regime type simultaneously, without any prior functional form restriction. Second, transition-specific and structural conditioning variables are added: a high inflation dummy; domestic credit to the private sector as a share of GDP, capturing financial sector development; capital account openness (Chinn-Ito index); a banking crisis dummy (Laeven and Valencia, 2020); a composite governance quality index (average of WGI government effectiveness, regulatory quality, and rule of law); and a dummy for EU membership or accession status. The inclusion of the EU variable is particularly important for the extended sample period, during which several transition economies completed accession while others began the process, creating a structural divergence in monetary and institutional conditions that the regression framework can only partially capture through fixed effects. Third, sub-period and regional group indicators are included to absorb systematic temporal and EU-status variation that would otherwise inflate the importance scores of substantive predictors.

### 3.4.3 Implementation

Random forest estimation is implemented in Stata using the rforest package (Schonlau and Zou, 2020). The dependent variable is GDP per capita growth. All forests use 500 trees; robustness to the number of trees and the predictor subset size is verified. Variable importance is reported as the mean decrease in prediction accuracy upon permutation, providing a comparability basis across subsamples.

## 3.5 Data

The sample consists of 27 transition economies of Central and Eastern Europe and the Commonwealth of Independent States over 1991–2019. Exchange rate regime data are taken from the Couharde and Grekou (2024) synthesis classification, which provides 671 usable country-year observations for the transition economy sample over this period. Growth regression variables — GDP per capita growth, initial GDP, government consumption, investment, inflation, terms of trade growth, population growth, population size, and trade openness — are drawn from the World Development Indicators. A high inflation dummy is constructed from the inflation series, taking the value one in country-years where annual inflation exceeds 40 percent. Capital account openness is measured by the Chinn-Ito index (Chinn and Ito, 2006). The composite governance quality index — the average of WGI government effectiveness, regulatory quality, and rule of law — is from the World Governance Indicators. Financial crisis episodes follow Laeven and Valencia (2020). EU membership and accession status is coded from official EU records. Variable names, descriptions and sources are further detailed in **Table A 1** and descriptive statistics in

**Table A 2** of the Appendix.

### 3.6. The Synthesis Classification: A Descriptive View

**Table 1** presents the distribution of Couharde-Grekou regime assignments across the transition economy sample over 1991–2019, distinguishing between EU members and non-EU members. Several patterns are worth noting. For the full sample, intermediate regimes are the modal arrangement throughout, accounting for 52 percent of country-year observations in the early transition period (1991–1999), before the distribution broadens as fixed and floating regimes each gain ground through the 2000s. Floating reaches its peak share in 2000–2007 (34.4 percent) and then contracts markedly, falling to 20.8 percent by 2014–2019, while the fixed regime share rises steadily from 16.4 percent in the early period to 37.7 percent in the latest sub-period.

The divergence between EU members and non-EU members is the most prominent structural feature of the table. EU-member economies follow a trajectory shaped by the gravitational pull of euro-area accession: their fixed regime share climbs from 21.0 percent in 1991–1999 to 55.4 percent in 2014–2019, as floating collapses from 32.1 to 7.7 percent over the same span — consistent with the currency board commitments and formal euro adoption of the newer member states. Non-EU economies display no such convergence; the intermediate arrangement remains dominant in every sub-period (ranging from 36.4 to 57.7 percent), the fixed share stays relatively flat at 25–33 percent, and floating is broadly stable at 28–31 percent across the four periods. The full-period distributions — Fixed: 35.3 versus 26.4 percent, Intermediate: 35.7 versus 43.7 percent, and Floating: 29.0 versus 29.9 percent for EU and non-EU members respectively — underscore that EU membership or accession status is a first-order determinant of the regime environment, a heterogeneity that the regression framework partially captures through the EU membership control variable and that the random forest analysis is specifically designed to exploit.

**Table 1. Exchange Rate Regime Distribution in Transition Economies, 1991–2019**

| Regime | 1991–1999 | 2000–2007 | 2008–2013 | 2014–2019 | Full period |
|---|---|---|---|---|---|
| **All transition economies** | | | | | |
| Fixed | 16.4% | 33.0% | 33.3% | 37.7% | 30.4% |
| Intermediate | 52.0% | 32.5% | 37.2% | 41.6% | 40.1% |
| Floating | 31.6% | 34.4% | 29.5% | 20.8% | 29.5% |
| N | 152 | 209 | 156 | 154 | 671 |
| **EU members** | | | | | |
| Fixed | 21.0% | 33.0% | 36.4% | 55.4% | 35.3% |
| Intermediate | 46.9% | 27.3% | 31.8% | 36.9% | 35.7% |
| Floating | 32.1% | 39.8% | 31.8% | 7.7% | 29.0% |
| N | 81 | 88 | 66 | 65 | 300 |
| **Non-EU members** | | | | | |
| Fixed | 11.3% | 33.1% | 31.1% | 24.7% | 26.4% |
| Intermediate | 57.7% | 36.4% | 41.1% | 44.9% | 43.7% |
| Floating | 31.0% | 30.6% | 27.8% | 30.3% | 29.9% |
| N | 71 | 121 | 90 | 89 | 371 |

*Source: Couharde-Grekou (2024) Synthesis Classification. EU members comprises the 11 transition economies that held EU membership by end of sample: Czech Republic, Estonia, Hungary, Latvia, Lithuania, Poland, Slovakia, Slovenia, Bulgaria, Romania, and Croatia. Non-EU members comprises the remaining 16 economies.*

**Figure 1** adds a dimension that **Table 1** cannot convey. While the categorical counts show which regime was most likely assigned in each period, the probability series reveal two further features. First, the degree of classification certainty — captured by the dashed certainty index — follows distinct trajectories across the two subgroups. EU members record the lowest certainty of any group in the early transition years (0.835 in 1991–1999), reflecting genuine ambiguity over the mixed peg-and-reform arrangements of the pre-accession period, before certainty rises sharply to 0.950 by 2014–2019 as euro-area commitments concentrate probability mass on fixed. Non-EU members, by contrast, display comparatively stable and high certainty throughout (0.886 in 1991–1999, remaining in the 0.880–0.910 range), suggesting that while their modal regime is intermediate, those assignments are themselves relatively unambiguous. Second, even within a given categorical assignment, the underlying probability mass is often well away from the corners of the simplex, particularly for EU members in the 2000–2007 period, when average probabilities across the three regime categories are nearly equal (0.341, 0.296, and 0.364 for fixed, intermediate, and floating respectively) — indicating that many observations carry genuine ambiguity across regime categories that a binary dummy simply discards. This variation in certainty is precisely what the probabilistic specification of Section 3.3 exploits. Average regime probabilities and classification certainty statistics are reported in **Table A 3** of the Appendix.

**Figure 1. Average exchange rate regime probabilities for transition economies, 1991–2019, with a certainty index overlay**

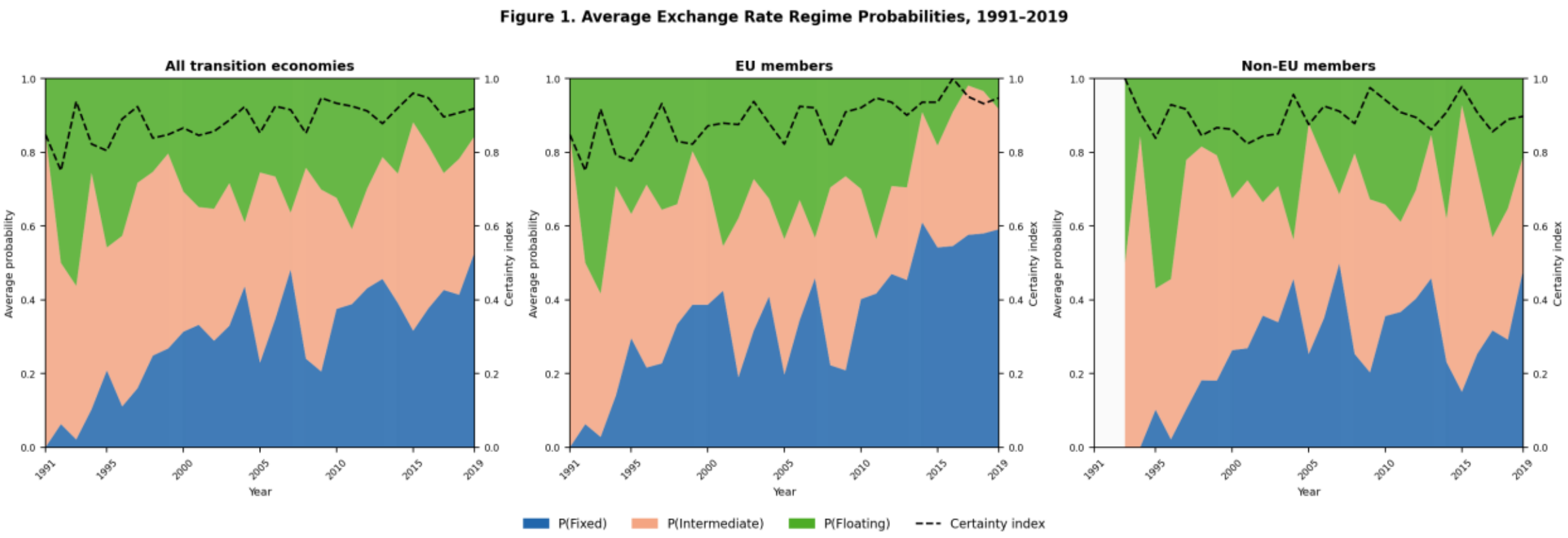

*Source: Couharde-Grekou (2024) Synthesis Classification.*

*Note: Stacked areas show average P(Fixed), P(Intermediate), and P(Floating) across countries in each subgroup by year. Dashed line shows the average certainty index (mean of the maximum probability across the three categories), scaled on the right axis (0–1).*

## 4. Regression Results

### 4.1 Baseline Fixed Effects: Categorical Exchange Rate Regimes

Across all three specifications, the growth penalty falls most heavily on economies operating under intermediate exchange rate arrangements — not on floaters, as much of the earlier developing-country literature would predict, but on the middle ground. The baseline fixed-effects estimate places the intermediate regime shortfall at roughly two percentage points per annum relative to fixed arrangements, a gap that narrows but survives the successive addition of sub-period dummies and country-time two-way controls. That the penalty persists through these increasingly demanding specifications matters: it implies the intermediate disadvantage is not simply a reflection of bad global years disproportionately affecting managed floaters, nor of country-specific characteristics correlated with regime choice, but something embedded in the regime arrangement itself.

Floating regimes also underperform the fixed benchmark in point-estimate terms, but the gap never achieves conventional significance. The hierarchy implied by the estimates — fixed first, floating second, intermediate last — sits uncomfortably with the standard debate, which has long been framed as a contest between pegs and floats. What the transition economy evidence instead suggests is that the real divide runs between regimes with clear institutional commitment on one side and discretionary management on the other. This reading finds natural support in Fisher's (2001) argument that intermediate arrangements combine the costs of both extremes: they surrender the credibility dividend of a hard anchor without acquiring the shock-absorption flexibility of a genuine float. In transition economies, where monetary credibility was scarce and institutional frameworks were being constructed in real time, that combination appears to have been particularly costly. The findings are consistent with De Grauwe and Schnabl (2004) for CEE economies and with Petreski (2014) for the broader transition context, and align with Cruz-Rodriguez (2022), who similarly found intermediate arrangements difficult to distinguish from floats in developing-country samples — though in our sample the intermediate shortfall relative to fixed is more precisely estimated than the floating one.

**Table 2** presents panel fixed-effects estimates of the benchmark growth regression (equation 1) with categorical exchange rate regime dummies, using the fixed regime as the reference category throughout. Three specifications are reported: country fixed effects alone (column 1), country fixed effects augmented with sub-period dummies to absorb common temporal shocks (column 2), and a two-way specification absorbing both country and year effects (column 3).

Across all three specifications, the growth penalty falls most heavily on economies operating under intermediate exchange rate arrangements — not on floaters, as much of the earlier developing-country literature would predict, but on the middle ground. The baseline fixed-effects estimate places the intermediate regime shortfall at roughly two percentage points per annum relative to fixed arrangements, a gap that narrows but survives the successive addition of sub-period dummies and country-time two-way controls. That the penalty persists through these increasingly demanding specifications matters: it implies the intermediate disadvantage is not simply a reflection of bad global years disproportionately affecting managed floaters, nor of country-specific characteristics correlated with regime choice, but something embedded in the regime arrangement itself.

Floating regimes also underperform the fixed benchmark in point-estimate terms, but the gap never achieves conventional significance. The hierarchy implied by the estimates — fixed first, floating second, intermediate last — sits uncomfortably with the standard debate, which has long been framed as a contest between pegs and floats. What the transition economy evidence instead suggests is that the real divide runs between regimes with clear institutional commitment on one side and discretionary management on the other. This reading finds natural support in Fisher's (2001) argument that intermediate arrangements combine the costs of both extremes: they surrender the credibility dividend of a hard anchor without acquiring the shock-absorption flexibility of a genuine float. In transition economies, where monetary credibility was scarce and institutional frameworks were being constructed in real time, that combination appears to have been particularly costly. The findings are consistent with De Grauwe and Schnabl (2004) for CEE economies and with Petreski (2014) for the broader transition context, and align with Cruz-Rodriguez (2022), who similarly found intermediate arrangements difficult to distinguish from floats in developing-country samples — though in our sample the intermediate shortfall relative to fixed is more precisely estimated than the floating one.

**Table 2. Baseline Fixed Effects**

| *Dependent variable: GDP per capita growth* | | | |
|---|---|---|---|
| | Country FE | Country + Period FE | Country + Year FE |
| | (1) | (2) | (3) |
| **Government consumption (% GDP)** | 0.405** | -0.115 | 0.052 |
| | [0.190] | [0.184] | [0.163] |
| **Gross fixed capital formation (% GDP)** | 0.291** | 0.092 | 0.214** |
| | [0.121] | [0.085] | [0.089] |
| **Terms of trade growth (%)** | -0.023 | -0.002 | -0.009 |
| | [0.021] | [0.020] | [0.019] |
| **Inflation, CPI (annual %)** | -0.039* | -0.05 | -0.049* |
| | [0.020] | [0.033] | [0.025] |
| **Population growth (%)** | -2.448*** | -1.846*** | -1.675*** |
| | [0.631] | [0.391] | [0.374] |
| **Log population** | 18.415* | 7.043 | 5.894 |
| | [9.782] | [5.539] | [4.485] |
| **Trade openness (% GDP)** | 0.071*** | 0.090*** | 0.070*** |
| | [0.018] | [0.029] | [0.022] |
| **ERR dummy: Floating** | -1.307 | -1.079 | -0.798 |
| | [1.013] | [0.646] | [0.630] |
| **ERR dummy: Intermediate** | -2.019** | -1.737*** | -0.944* |
| | [0.815] | [0.566] | [0.512] |
| | | | |
| **Observations** | 416 | 416 | 414 |
| **R-squared** | 0.164 | 0.429 | 0.66 |
| **Number of cid** | 24 | 24 | |
| **Country FE** | Yes | Yes | Yes |
| **Year FE** | No | No | Yes |
| **Period FE** | No | Yes | No |

*Source: Author’s calculations. *, ** and *** refer to a statistical significance at the 10%, 5% and 1% level, respectively. Robust standard errors provided in brackets.*

Among the control variables, coefficients conform to theoretical priors. Investment (gross fixed capital formation) enters positively and significantly in the baseline and two-way specifications, consistent with the neoclassical and endogenous growth channels documented by Barro and Sala-i-Martin (2004). Trade openness is positive and significant across all columns, reflecting the growth dividend from international market integration that characterized successful transition (Sachs, 1996). Population growth carries large, consistently negative and significant coefficients. Inflation exerts a negative and marginally significant effect in two of three specifications. Model fit rises from $R^2 = 0.164$ in the country-FE baseline to 0.429 with period controls and 0.660 in the two-way specification.

### 4.2 Probabilistic Exchange Rate Regime Specification

makes fuller use of the Couharde-Grekou synthesis classification by replacing hard dummies with the underlying probability series. Rather than asking whether a country-year was assigned to a given regime, this specification asks how much probability mass it carries toward each arrangement — treating classification uncertainty as usable information rather than noise to be discarded. The implicit benchmark remains P(Fixed): what the coefficients on P(Floating) and P(Intermediate) capture is the growth consequence of probability mass shifting away from fixity and toward the other two poles.

The story that emerges reinforces and sharpens the **Table 2** findings. The intermediate penalty survives across all four specifications and is, if anything, more precisely estimated: the point estimates are consistently larger in magnitude than their categorical counterparts, reflecting the fact that the continuous probability measure extracts signal that hard classification leaves on the table. A country-year sitting with high certainty in an intermediate arrangement fares worse than one ambiguously straddling fixed and intermediate — the probability specification captures that gradient, and it penalizes it. Floating, again, underperforms fixed in point-estimate terms, but only weakly and intermittently in significance terms, replicating the pattern from **Table 2**.

The most substantively interesting test is column 4, which adds the certainty index to ask a sharper question: is it the regime that matters, or merely the clarity with which it is identified? The answer is unambiguous. The certainty index enters insignificantly, while the regime probability coefficients are essentially unchanged in magnitude and significance. Growth responds to the nature of the exchange rate arrangement, not to how confidently it can be labelled. This finding resonates with Calvo and Reinhart's (2002) fear-of-floating framework: what drives outcomes is the actual exchange rate behavior implied by higher fixity probability — the discipline, credibility, and reduced currency risk that accompany a genuine commitment to limited flexibility — not the statistical confidence of the observer classifying it.

**Table 3** makes fuller use of the Couharde-Grekou synthesis classification by replacing hard dummies with the underlying probability series. Rather than asking whether a country-year was assigned to a given regime, this specification asks how much probability mass it carries toward each arrangement — treating classification uncertainty as usable information rather than noise to be discarded. The implicit benchmark remains P(Fixed): what the coefficients on P(Floating) and P(Intermediate) capture is the growth consequence of probability mass shifting away from fixity and toward the other two poles.

The story that emerges reinforces and sharpens the **Table 2** findings. The intermediate penalty survives across all four specifications and is, if anything, more precisely estimated: the point estimates are consistently larger in magnitude than their categorical counterparts, reflecting the fact that the continuous probability measure extracts signal that hard classification leaves on the table. A country-year sitting with high certainty in an intermediate arrangement fares worse than one ambiguously straddling fixed and intermediate — the probability specification captures that gradient, and it penalizes it. Floating, again, underperforms fixed in point-estimate terms, but only weakly and intermittently in significance terms, replicating the pattern from **Table 2**.

The most substantively interesting test is column 4, which adds the certainty index to ask a sharper question: is it the regime that matters, or merely the clarity with which it is identified? The answer is unambiguous. The certainty index enters insignificantly, while the regime probability coefficients are essentially unchanged in magnitude and significance. Growth responds to the nature of the exchange rate arrangement, not to how confidently it can be labelled. This finding resonates with Calvo and Reinhart's (2002) fear-of-floating framework: what drives outcomes is the actual exchange rate behavior implied by higher fixity probability — the discipline, credibility, and reduced currency risk that accompany a genuine commitment to limited flexibility — not the statistical confidence of the observer classifying it.

**Table 3. Results from the Probabilistic Specification**

| | *Dependent variable: GDP per capita growth* | | | |
|---|---|---|---|---|
| | Country FE | Country + Period FE | Country + Year FE | Country FE + Certainty control |
| | (1) | (2) | (3) | (4) |
| **Government consumption (% GDP)** | 0.397** | -0.121 | 0.041 | 0.395** |
| | [0.187] | [0.181] | [0.163] | [0.187] |
| **Gross fixed capital formation (% GDP)** | 0.291** | 0.095 | 0.214** | 0.289** |
| | [0.121] | [0.084] | [0.090] | [0.120] |
| **Terms of trade growth (%)** | -0.023 | -0.003 | -0.009 | -0.024 |
| | [0.022] | [0.019] | [0.018] | [0.023] |
| **Inflation, CPI (annual %)** | -0.033* | -0.043 | -0.046* | -0.033* |
| | [0.019] | [0.031] | [0.024] | [0.019] |
| **Population growth (%)** | -2.367*** | -1.784*** | -1.639*** | -2.352*** |
| | [0.595] | [0.357] | [0.371] | [0.597] |
| **Log population** | 17.600* | 6.07 | 5.5 | 17.731* |
| | [9.645] | [5.778] | [4.357] | [9.608] |
| **Trade openness (% GDP)** | 0.067*** | 0.086*** | 0.069*** | 0.068*** |
| | [0.016] | [0.027] | [0.021] | [0.017] |
| **P(Floating) — CC synthesis** | -1.684 | -1.710** | -1.12 | -1.657 |
| | [1.215] | [0.731] | [0.725] | [1.220] |
| **P(Intermediate) — CC synthesis** | -2.726** | -2.601*** | -1.400** | -2.744** |
| | [1.076] | [0.749] | [0.633] | [1.094] |
| | | | | -0.585 |
| **Classification certainty (max prob)** | | | | [1.121] |
| | | | | |
| **Observations** | 416 | 416 | 414 | 416 |
| **R-squared** | 0.173 | 0.44 | 0.663 | 0.173 |
| **Number of cid** | 24 | 24 | 24 | 24 |
| **Country FE** | Yes | Yes | Yes | Yes |
| **Year FE** | No | No | Yes | No |
| **Period FE** | No | Yes | No | No |
| **Certainty ctrl** | No | No | No | Yes |

*Source: Author's calculations. *, ** and *** refer to a statistical significance at the 10%, 5% and 1% level, respectively. Robust standard errors provided in brackets.*

### 4.3 System GMM: Dynamic Panel Specification

The fixed-effects estimator in **Table 2** and **Table 3** faces a familiar challenge: regime choices are not random. Countries experiencing growth difficulties may abandon fixed arrangements, biasing the FE comparison in favor of floats and against pegs. **Table 4** addresses this directly through system GMM, instrumenting the regime variables with their own lags to break the contemporaneous feedback between growth outcomes and regime assignment. Three specifications are reported: categorical ERR (column 1), probabilistic ERR (column 2), and an augmented probabilistic specification controlling for banking crises, composite governance quality (average of WGI government effectiveness, regulatory quality, and rule of law), and EU membership (column 3). The diagnostic tests offer no grounds for concern — second-order serial correlation and instrument validity are both comfortably supported across all three columns — and the lagged dependent variable enters positive and highly significant throughout, confirming the dynamic persistence of growth that motivates the GMM framework in the first place.

The central finding is that correcting for endogeneity does not attenuate the intermediate regime penalty — it dramatically amplifies it. Where the two-way fixed-effects estimate placed the intermediate shortfall at roughly one to two percentage points annually, the GMM estimate puts it at seven to ten percentage points. This is not a statistical artefact but an informative signal about the direction of bias: fixed regimes were disproportionately abandoned during episodes of growth distress, which caused the FE estimator to understate their advantage. Once that reverse causality is instrumented out, the fixed regime premium asserts itself with considerable force — a pattern consistent with Levy-Yeyati and Sturzenegger (2003) and Petreski (2014), both of whom found that addressing endogeneity moves estimates in the same direction. Floating regimes continue to underperform the fixed benchmark in point-estimate terms, but without achieving significance, preserving the same hierarchy established in the earlier tables.

The augmented specification adds banking crisis exposure, institutional quality, and EU membership status to test whether the ERR effect is merely proxying for these correlated institutional dimensions. None of the three additional controls reaches conventional significance, and the intermediate penalty is essentially unchanged. The exchange rate regime effect has its own independent bite — it is not a stand-in for better governance, crisis vulnerability, or the particular trajectory of EU-accession economies.

**Table 4. Results from the System-GMM estimation**

| *Dependent variable: GDP per capita growth* | | | |
|---|---|---|---|
| | ERR as categorical | ERR as probabilistic | ERR probabilistic + controls |
| | (1) | (2) | (3) |
| **GDP per capita growth (%, WDI) = L,** | 0.501*** | 0.435*** | 0.326*** |
| | [0.129] | [0.102] | [0.113] |
| **Log GDP per capita 1990 (convergence)** | 0.523 | 0.367 | 1.107 |
| | [1.226] | [0.919] | [1.186] |
| **Government consumption (% GDP)** | 0.131 | 0.243 | 0.211 |
| | [0.247] | [0.184] | [0.179] |
| **Gross fixed capital formation (% GDP)** | -0.034 | 0.074 | 0.2 |
| | [0.221] | [0.159] | [0.170] |
| **Terms of trade growth (%)** | -0.015 | -0.016 | -0.031 |
| | [0.037] | [0.031] | [0.031] |
| **Inflation, CPI (annual %)** | 0.043* | 0.038** | 0.039** |
| | [0.023] | [0.018] | [0.015] |
| **Population growth (%)** | -0.003 | -0.226 | -0.652 |
| | [0.857] | [0.707] | [0.745] |
| **Log population** | -0.488 | -0.424 | -0.142 |
| | [0.680] | [0.427] | [0.539] |
| **Trade openness (% GDP)** | -0.087** | -0.062** | -0.028 |
| | [0.034] | [0.027] | [0.038] |
| **ERR: Floating** | -3.486 | -0.533 | -0.524 |
| | [3.069] | [3.043] | [2.864] |
| **ERR: Intermediate** | -10.419*** | -7.973*** | -7.573*** |
| | [2.584] | [2.736] | [2.662] |
| **Banking crisis dummy (L&V 2020)** | | | -1.859* |
| | | | [0.938] |
| **WGI composite (avg GE + RQ + RL)** | | | -0.071 |
| | | | [0.066] |
| **EU membership dummy** | | | -0.935 |
| | | | [1.345] |
| | | | |
| **Observations** | 416 | 416 | 416 |
| **Number of cid** | 24 | 24 | 24 |
| **AR2 p** | 0.798 | 0.753 | 0.574 |
| **Hansen J p** | 0.455 | 0.326 | 0.52 |

*Source: Author's calculations. *, ** and *** refer to a statistical significance at the 10%, 5% and 1% level, respectively. Robust standard errors provided in brackets.*

### 4.4 Temporal Heterogeneity and Regional Subgroups

**Table 5** asks a more pointed question: does the regime-growth relationship hold uniformly across the transition experience, or is it concentrated in particular periods and country environments? The answer is emphatically the latter, and the pattern of heterogeneity is itself informative.

The sub-period results reveal a sharp temporal gradient. In the years before 2003 — spanning early liberalization, the first wave of stabilization programs, and the late-1990s crisis episodes — the growth gap between regime types is at its widest. Both floating and intermediate arrangements significantly underperformed the fixed benchmark in this period, with floating carrying the larger penalty. This is precisely the context where the credibility channel, emphasized by De Grauwe and Schnabl (2004) and Ghosh et al. (1997), would be expected to dominate: nominal anchors were scarce, inflationary legacies from price liberalization were still being absorbed, and floating economies lacked the institutional and financial infrastructure needed to convert exchange rate flexibility into macroeconomic stability. Fixed arrangements, by contrast, provided a hard commitment that substituted for the credibility central banks had not yet earned. Moving forward into the 2003–2008 and 2009–2019 sub-periods, all regime differentials fade to statistical insignificance. Floating economies even post positive point estimates in both later sub-periods, and intermediate penalties become negligible. The implication is that as institutional capacity converged, financial systems deepened, and EU-integrated economies moved toward common monetary frameworks — as documented by Belhocine et al. (2016) — the exchange rate regime ceased to be the binding constraint on growth that it had been in the volatile early years.

The EU versus non-EU breakdown tells the same story from a cross-sectional angle. Among EU-member economies, no regime variable achieves significance. The institutional anchor of EU accession — with its accompanying requirements on monetary frameworks, fiscal governance, and financial regulation — appears to have rendered the specific exchange rate arrangement largely redundant as a growth determinant. Among non-EU transition economies, by contrast, the intermediate regime penalty remains large, negative, and statistically significant. For these countries — primarily CIS economies operating with weak monetary institutions and without the external discipline of accession conditionality — the choice between genuine commitment and discretionary management continues to carry real growth consequences. This contrast maps directly onto the conditionality identified by Slavtcheva (2015), who found fixed regime growth advantages concentrated in countries with limited financial development, and by Khan, Kebewar, and Nenovsky (2013), who showed that monetary regime consequences depend critically on the institutional environment in which they operate. Exchange rate regime choices matter most precisely where institutions are weakest — and, by extension, least for those economies that have already resolved the credibility problem through accession.

**Table 5. Temporal and EU Membership Heterogeneity**

| | **Periods** | | | **EU membership** | |
|---|---|---|---|---|---|
| | Before 2003 | 2003-2008 | 2009-2019 | EU members | Non-EU members |
| | (1) | (2) | (3) | (4) | (5) |
| **ERR: Floating** | -3.415*** | 1.64 | 3.047 | 3.604 | -2.168 |
| | [1.166] | [2.339] | [4.126] | [4.344] | [2.052] |

| ERR: Intermediate | -2.863* | -1.104 | -0.127 | -2.804 | -6.428** |
|---|---|---|---|---|---|
| | [1.534] | [2.756] | [2.294] | [3.394] | [2.906] |
| | | | | | |
| Observations | 57 | 117 | 242 | 161 | 255 |
| Number of cid | 19 | 20 | 24 | 11 | 24 |

*Source: Author's calculations. *, ** and *** refer to a statistical significance at the 10%, 5% and 1% level, respectively. Robust standard errors provided in brackets. Other controls included but not presented due to space.*

## 5. Random Forest Results

### 5.1 Model Fit

**Table 6** reports the out-of-bag (OOB) prediction errors from the seven random forest models estimated in this paper. The OOB error — the mean squared prediction error on observations withheld from each bootstrap sample — provides an approximately unbiased estimate of out-of-sample prediction accuracy (Breiman, 2001; Schonlau and Zou, 2020).

The pooled forest achieves an OOB error of 2.036, confirming genuine out-of-sample predictive content across the full sample. Regime-specific forests display heterogeneous fit that is itself informative. The fixed-regime subsample is the most predictable (OOB = 1.77), followed by the floating subsample (OOB = 2.011). The intermediate subsample is markedly harder to predict (OOB = 2.561), consistent with the conceptual heterogeneity of this category: managed floats, crawling bands, and ambiguous pegs share a classification label but represent fundamentally different macroeconomic dynamics across diverse country-time contexts. The lower OOB error for fixed-regime observations implies that growth dynamics within hard peg arrangements are more structured and replicable — the predictor set explains a larger share of growth variation when the nominal anchor channels adjustment through a more consistent set of determinants. This resonates with Husain et al.'s (2005) finding that regime durability — highest for hard pegs — is associated with more predictable macroeconomic outcomes.

Among EU-membership forests, the EU subsample achieves lower OOB error (2.003), reflecting the greater homogeneity of growth dynamics in a group of EU economies having converged toward common institutional frameworks. Non-EU economies are more difficult to predict (OOB = 2.117), consistent with the structural diversity of this group spanning resource-rich and resource-poor economies, a wide range of governance quality, and monetary strategies from currency boards to near-hyperinflation episodes.

**Table 6. Out-of-Bag Prediction Errors — Random Forest Models**

| Forest | OOB error |
|---|---|
| **Pooled** | **2.036** |
| By regime type | |
| Fixed | 1.770 |
| Floating | 2.011 |
| Intermediate | 2.561 |
| By EU-member status | |

| EU | 2.003 |
|---|---|
| Non-EU | 2.117 |

*Source: Author's calculations. OOB error is the mean squared prediction error on observations withheld from each bootstrap sample.*

### 5.2 Variable Importance: Pooled Forest

**Figure 2** presents the all predictors ranked by permutation importance from the pooled forest. The permutation importance score measures the mean decrease in prediction accuracy when each predictor is randomly permuted across all 500 trees, integrating both direct effects and interaction pathways without any prior functional form restriction (Breiman, 2001).

**Figure 2. Variable Importance – Pooled Forest**

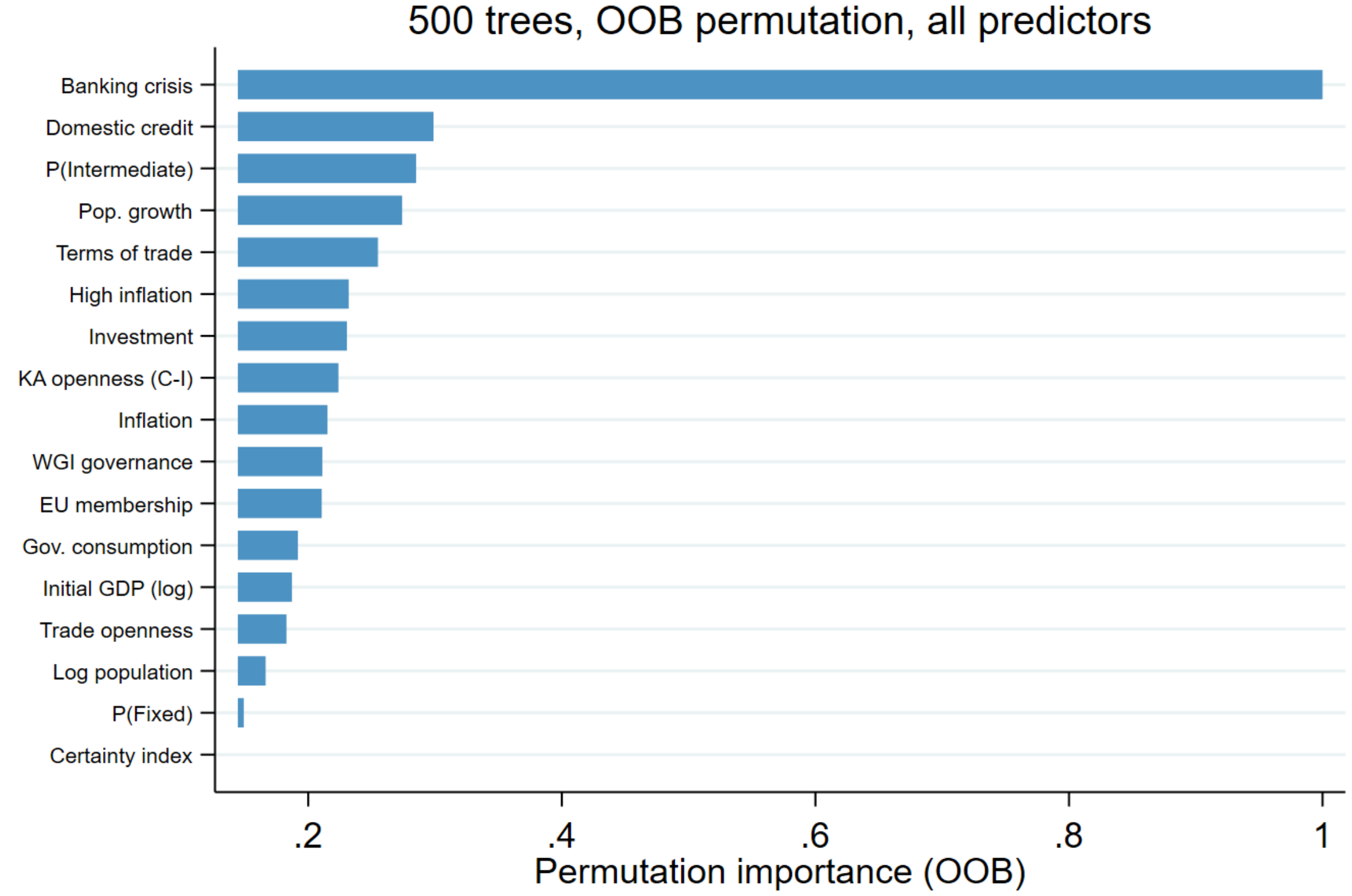

*Source: Author's estimations.*

Banking crises stand apart from every other predictor in the dataset. Their importance score dwarfs the second-ranked variable by a factor of more than three, a gap so large that it dominates the visual scale and compresses all other bars into apparent similarity. This is not a statistical artefact — OOB permutation importance measures the damage to out-of-bag prediction accuracy from randomly shuffling a variable, and shuffling banking crisis status destroys predictive accuracy in a way that nothing else comes close to matching. The transition economy sample lived through multiple systemic banking events, and the forest identifies crisis years as the single most informative feature of the growth environment, operating through non-linear pathways that the GMM specifications could not fully capture.

Below that outlier, the remaining predictors cluster in a band between roughly 0.10 and 0.30. Domestic credit ranks second, reflecting the importance of financial development depth in these structurally transforming economies. Crucially, intermediate ERR ranks third — the highest-ranked ERR variable and the only one that registers meaningfully in the pooled forest. This directly corroborates the regression hierarchy: it is variation in intermediate regime assignment that carries information about growth outcomes, not fixed or floating per se. Fixed ERR sits near the bottom of the distribution and the certainty index registers essentially zero importance, reinforcing the Table 3 result that regime type rather than classification clarity drives growth.

Population growth, terms of trade, high inflation, and investment cluster in positions four through seven, broadly consistent with the significant regression controls. The standard convergence variable — initial GDP — falls surprisingly low, likely reflecting that the non-linear tree structure captures convergence dynamics through interactions with the financial and institutional variables rather than as a standalone predictor.

### 5.3 Regime-Specific Variable Importance

**Figure 3** compares variable importance rankings across the three regime-specific forests. This comparison constitutes the primary methodological contribution of the random forest analysis: unlike the regression framework, which can only assess regime heterogeneity through pre-specified interaction terms, the forests estimated separately for fixed, intermediate, and floating subsamples reveal which factors most powerfully predict growth within each regime context without any prior theoretical commitment.

**Figure 3. Variable Importance by Regime Type**

*Source: Author's estimations.*

Banking crisis tops all three forests, but the gap between regime types is informative: its importance is roughly equal and near-maximum under fixed and intermediate regimes, but noticeably lower under floating. This is consistent with the idea that floating economies can partially absorb crisis shocks through exchange rate adjustment, reducing the discrete predictive bite of crisis episodes.

The most striking single feature of the figure is the terms of trade bar for floating regimes, which reaches the maximum scaled value — equal to banking crisis — while remaining relatively modest under fixed and near-trivial under intermediate. The economic logic is compelling: floating exchange rates are the primary mechanism through which terms-of-trade shocks are transmitted into domestic real income, so ToT variation is the dominant feature of the floating growth environment. Fixed-regime economies, by contrast, absorb external shocks through reserves, fiscal adjustment, and domestic price flexibility rather than through the exchange rate, reducing the forest's need to exploit ToT variation.

Inflation ranks as the third most important predictor under fixed regimes but is the weakest predictor of all under intermediate — the lowest bar in the intermediate forest. Maintaining a peg requires price discipline, so inflation variation is highly informative about whether a fixed arrangement is sustainable and whether its growth dividend materializes. Under intermediate regimes, the managed float allows some exchange rate response to inflationary pressure, attenuating the predictive signal. Domestic credit follows a similar pattern: very high importance under fixed (rank 2) and floating (moderate), but low under intermediate. The intermediate forest overall returns the smallest importance values across virtually every predictor, suggesting that growth under managed arrangements is the hardest to predict from standard macroeconomic variables — or equivalently, that intermediate-regime growth is the most idiosyncratic, driven by discretionary policy responses that the forest's predictors do not well capture.

EU membership shows an interesting reversal: it is more important for floating-regime growth than for fixed. This likely reflects that EU accession timing interacts with the exchange rate flexibility choice in ways that matter for growth trajectories of floaters, while for fixed-regime countries the EU anchor is already embedded in the commitment structure.

**Figure 4** compares variable importance rankings across the EU-membership forests and tells perhaps the most coherent institutional story of the three. With only one exception, non-EU importance values exceed EU values for every predictor in the dataset — in several cases by a factor of two or more. This asymmetry is the figure's central message: the standard macroeconomic predictors in this model describe the non-EU growth environment far better than the EU one.

Banking crisis again anchors both forests at the top with equal importance — financial crises are universally destructive regardless of institutional integration, and no amount of EU convergence insulates an economy from a systemic banking event. Below that shared peak, however, the two forests diverge in ways that speak directly to the paper's central argument. The most striking feature is the position of intermediate ERR in the non-EU forest, where it ranks second overall with an importance score of approximately 0.57 — by far the highest ERR signal in any of the three figures, and nearly matching the banking crisis bar in relative terms. The exchange rate regime is not merely one of many conditioning factors for non-EU transition economies: it is the dominant non-crisis predictor of growth outcomes, with the intermediate arrangement carrying more predictive weight than domestic credit, population dynamics, investment, or any other standard control. This directly reinforces the regression finding that the intermediate penalty is sharpest and most precisely estimated in the non-EU subsample.

Among EU members, intermediate ERR still registers at around 0.24, ranking sixth, but the gap relative to non-EU is large and telling. The accession framework appears to have partially domesticated the intermediate regime disadvantage by embedding external monetary discipline even for countries that did not formally adopt the euro, reducing the predictive bite of regime variation without fully eliminating it. Fixed ERR and the certainty index show a mild reversal of the general pattern: their importance is slightly higher for EU members than for Non-EU members, suggesting that within the EU group, the degree of fixity commitment and the clarity of regime assignment carry marginally more signal — consistent with the heterogeneous euro-area accession paths of the newer member states, where the transition from candidate to full monetary union membership is precisely the dimension along which fixed-regime growth effects concentrate.

**Figure 4. Variable Importance by EU Membership Status**

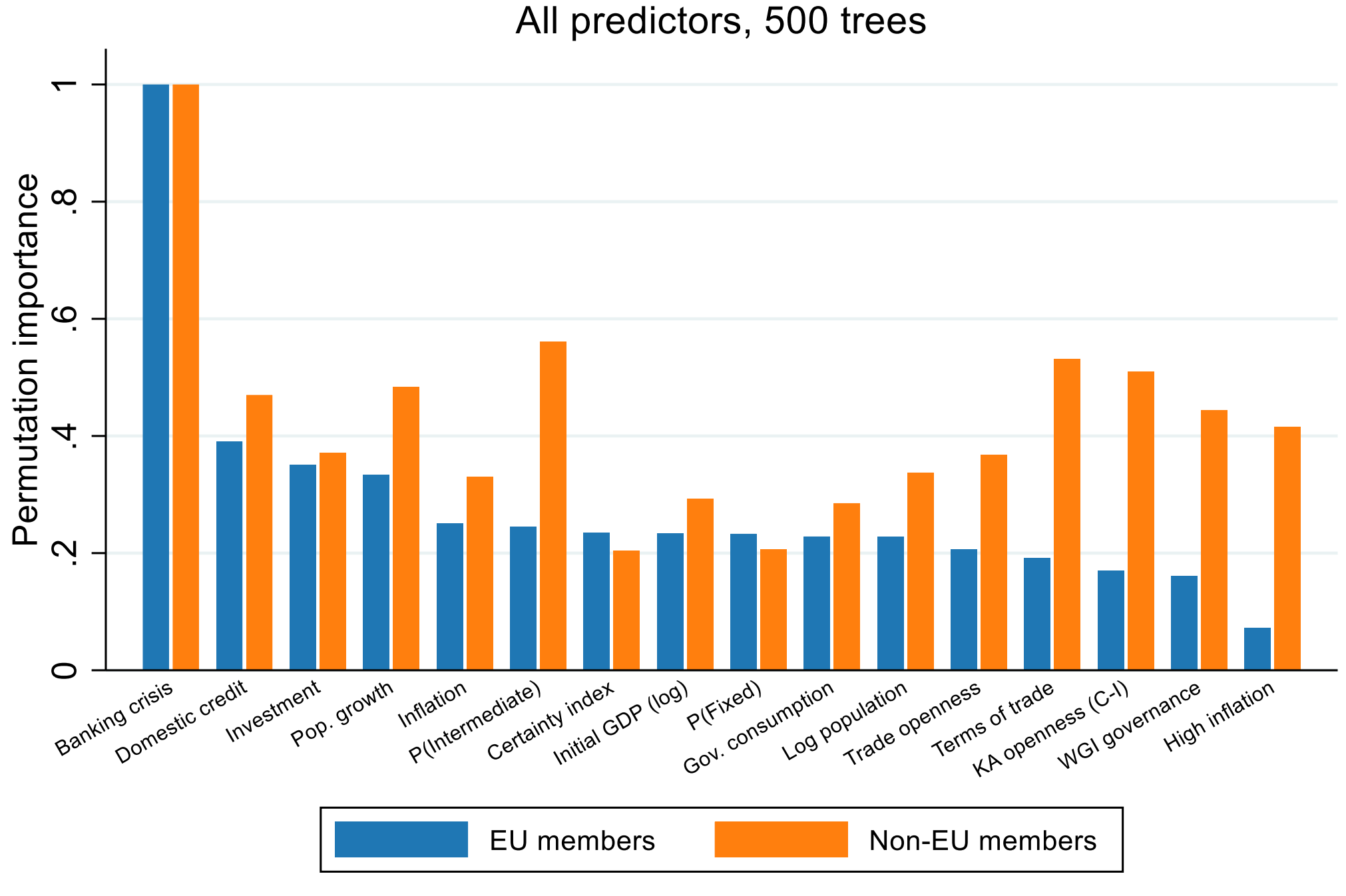

*Source: Author's estimations.*

The rest of the figure reproduces and quantifies the asymmetry observed before: terms of trade, KA openness, WGI governance, and high inflation all register substantially higher importance for non-EU economies. High inflation in particular falls to near-zero for EU members — the institutional anchor of accession has rendered it a non-event in that sample — while remaining a significant growth predictor for non-EU economies. The overall picture is one of a fundamentally different growth environment: non-EU transition economies operate in a space where macroeconomic fundamentals, external conditions, and — centrally — the exchange rate regime jointly determine outcomes in ways the forest can exploit, while EU-integrated economies have partially substituted institutional convergence for the domestic policy variation that makes these predictors informative.

### 6. Conclusion

This paper has revisited the relationship between exchange rate regimes and economic growth for 27 transition economies over 1991–2019. Three innovations distinguish it from earlier work: the use of the Couharde-Grekou (2024) probabilistic synthesis classification, which treats regime uncertainty as usable information rather than noise to be discarded; an extended sample that captures the Global Financial Crisis, post-crisis divergence, and the regional turbulence of 2014–15; and random forest estimation that identifies the growth-conditioning environment within each regime type and institutional group without imposing functional form or pre-specifying interactions.

The central finding is robust and unambiguous: intermediate exchange rate regimes are the worst-performing arrangement across every estimator and specification. The growth penalty relative to fixed regimes ranges from approximately one percentage point under two-way fixed effects to ten percentage points under system GMM, where correcting for reverse causality — countries abandoning fixed arrangements during downturns — strengthens rather than attenuates the fixed-regime advantage. Floating regimes also underperform fixed in sign throughout, but that gap is not robustly identified once endogeneity is addressed. The probabilistic specification confirms that what drives these differentials is the nature of the exchange rate arrangement, not the certainty with which it is classified — regime type matters, classification ambiguity does not. This ordering is consistent with the view that intermediate regimes in institutionally thin transition environments combined the costs of both extremes: the fiscal and monetary constraints of a commitment without its credibility dividend, and the external vulnerability of openness without its shock-absorption flexibility.

The regime effect is, however, sharply conditional. It is concentrated in the pre-2003 period, when exchange rate anchoring was the primary available instrument for establishing monetary credibility, and it is driven entirely by non-EU transition economies. Among EU members, no significant regime effect is detectable at any horizon — institutional convergence under accession has rendered the exchange rate choice largely inconsequential for growth. Among non-EU economies, the intermediate penalty reaches −6.4 percentage points under GMM. The descriptive evidence reinforces this divide structurally: EU-member economies have converged strongly toward fixed arrangements over the sample period, with the fixed-regime share rising from 21 to 55 percent by 2014–2019, while non-EU economies maintain broadly stable and distributed regime shares throughout.

The random forest analysis provides non-parametric confirmation of this picture while adding texture. In the pooled forest, banking crises dominate prediction by a wide margin, with intermediate the only ERR variable registering a meaningful signal — directly mirroring the regression hierarchy. The regime-conditioned forests reveal structurally distinct growth environments: terms-of-trade variation dominates under floating, consistent with the exchange rate's shock-absorption role; inflation discipline and financial depth lead under fixed, consistent with the credibility channel; and the intermediate subsample returns the weakest and most diffuse importance profile across all predictors, a non-parametric reflection of the regression's main finding. The EU/non-EU forests sharpen the institutional story further: intermediate ERR ranks second in the non-EU forest with an importance score approaching 0.6 — the strongest ERR signal in any decomposition — while registering more modestly for EU members. High inflation, governance quality, and external openness variables all show substantially higher predictive importance in the non-EU forest, confirming that EU integration has substituted external institutional anchors for the domestic macroeconomic variation that makes these predictors informative.

The policy implications are concentrated where the evidence points. For non-EU transition economies, the exchange rate regime retains real growth consequences, and the consistent finding that intermediate arrangements are the worst-performing option points toward a genuine corner solution: a hard peg with adequate institutional backing or a credible inflation-targeting float, rather than the discretionary middle ground the evidence identifies as most costly. For EU-integrated economies, the convergence process has already rendered the question largely moot — the growth consequences of eventual euro adoption are likely modest relative to the institutional forces that have already eroded the regime-growth nexus within this group.

Two caveats and two directions for future research round out the assessment. The GMM results rely on instrument validity that Hansen J statistics support but cannot establish with certainty, and the random forest importance rankings identify predictive relevance rather than structural causation. On the research frontier, the interaction between exchange rate regime and labor market flexibility identified by Kuokštis et al. (2025) merits direct investigation in the transition context — the post-2003 attenuation of regime effects may partly reflect institutional labor market reforms rather than monetary convergence alone. And extending the probabilistic classification framework to the post-2019 period, during which transition economies faced the COVID-19 shock, the energy crisis, and the spillovers from the war in Ukraine, would test whether the credibility channel identified for the early transition period has reasserted itself under renewed exceptional conditions.

**Appendix**

**Table A 1. Variables' names, descriptions and sources**

| Variable | Description | Source |
|---|---|---|
| **Dependent variable** | | |
| GDP per capita growth | Annual growth rate of GDP per capita (%) | *World Development Indicators* |
| **Growth regression controls** | | |
| Initial GDP (log) | Log of GDP per capita in 1990; captures conditional convergence | *World Development Indicators* |
| Government consumption | General government final consumption expenditure (% of GDP) | *World Development Indicators* |
| Investment | Gross fixed capital formation (% of GDP) | *World Development Indicators* |
| Inflation | Annual consumer price inflation (%) | *World Development Indicators* |
| Terms of trade growth | Annual growth rate of the terms of trade index (%) | *World Development Indicators* |
| Population growth | Annual population growth rate (%) | *World Development Indicators* |
| Population size (log) | Log of total population | *World Development Indicators* |
| Trade openness | Exports plus imports as a share of GDP (%) | *World Development Indicators* |
| **Exchange rate regime variables** | | |
| P(Fixed) | Posterior probability of fixed exchange rate regime assignment | *Couharde and Grekou (2024)* |
| P(Intermediate) | Posterior probability of intermediate exchange rate regime assignment | *Couharde and Grekou (2024)* |
| P(Floating) | Posterior probability of floating exchange rate regime assignment; reference in probabilistic regression | *Couharde and Grekou (2024)* |
| Fixed regime dummy | Binary; =1 if SC3w modal assignment is fixed | *Couharde and Grekou (2024)* |
| Intermediate regime dummy | Binary; =1 if SC3w modal assignment is intermediate | *Couharde and Grekou (2024)* |
| Floating regime dummy | Binary; =1 if SC3w modal assignment is floating; reference category in regressions | *Couharde and Grekou (2024)* |
| Certainty index | Max(P(Fixed), P(Intermediate), P(Floating)); measures classification certainty | *Derived from Couharde and Grekou (2024)* |
| **Additional random forest predictors** | | |
| Domestic credit | Domestic credit to private sector (% of GDP) | *World Development Indicators* |
| Capital account openness | Chinn-Ito KAOPEN index; higher values indicate greater financial openness | *Chinn and Ito (2006)* |
| Governance quality | Average of WGI government effectiveness, regulatory quality, and rule of law | *World Governance Indicators* |
| Banking crisis | Binary; =1 in years of systemic banking crisis | *Laeven and Valencia (2020)* |
| High inflation dummy | Binary; =1 if annual inflation exceeds 40% | *Constructed from World Development Indicators* |
| EU membership | Binary; =1 if country holds EU membership or active accession status | *Official EU records* |
| Sub-period | Categorical: 1=1991–1997, 2=1998–2002, 3=2003–2008, 4=2009–2019 | *Constructed* |

**Table A 2. Descriptive statistics**

| Variable | Observations | Mean | Standard deviation | Minimum | Maximum |
|---|---|---|---|---|---|
| GDP per capita growth | 416 | 4.186 | 4.198 | -14.781 | 14.62 |
| Initial GDP (log) | 416 | 8.494 | 0.713 | 6.674 | 9.519 |
| Government consumption | 416 | 18.045 | 3.853 | 9.526 | 33.316 |
| Investment | 416 | 23.056 | 4.873 | 13.226 | 38.717 |
| Terms of trade growth | 416 | 101.872 | 11.856 | 63.044 | 157 |
| Inflation | 416 | 5.975 | 11.026 | -1.584 | 168.62 |
| Population growth | 416 | -0.323 | 0.792 | -3.848 | 2.558 |
| Population size (log) | 416 | 15.73 | 1.183 | 14.089 | 18.803 |
| Trade openness | 416 | 105.066 | 32.721 | 46.287 | 188.733 |
| Fixed regime dummy | 416 | 0.317 | 0.466 | 0 | 1 |
| Intermediate regime dummy | 416 | 0.377 | 0.485 | 0 | 1 |
| Floating regime dummy | 416 | 0.298 | 0.458 | 0 | 1 |
| P(Fixed) | 416 | 0.341 | 0.426 | 0 | 1 |
| P(Intermediate) | 416 | 0.368 | 0.417 | 0 | 1 |
| P(Floating) | 416 | 0.291 | 0.42 | 0 | 1 |
| Certainty index | 416 | 0.902 | 0.153 | 0.458 | 1 |
| Domestic credit | 298 | 45.017 | 17.371 | 7.125 | 101.383 |
| Capital account openness | 416 | 0.633 | 1.447 | -1.939 | 2.281 |
| Governance quality | 416 | 56.004 | 11.314 | 31.286 | 78.89 |
| Banking crisis | 416 | 0.07 | 0.255 | 0 | 1 |
| EU membership | 416 | 0.387 | 0.488 | 0 | 1 |
| High inflation dummy | 416 | 0.017 | 0.129 | 0 | 1 |

**Table A 3. Exchange Rate Regime Distribution in Transition Economies, 1991–2019: probabilities and classification certainty**

| Group / Sub-period | P(Fixed) | P(Intermediate) | P(Floating) | Certainty index | N |
|---|---|---|---|---|---|
| **All transition economies** | | | | | |
| 1991–1999 | 0.173 | 0.501 | 0.326 | 0.859 | 149 |
| 2000–2007 | 0.344 | 0.335 | 0.321 | 0.883 | 211 |
| 2008–2013 | 0.349 | 0.353 | 0.298 | 0.908 | 156 |
| 2014–2019 | 0.408 | 0.393 | 0.198 | 0.925 | 156 |
| Full period | 0.322 | 0.389 | 0.288 | 0.893 | 672 |
| **EU members** | | | | | |
| 1991–1999 | 0.228 | 0.440 | 0.332 | 0.835 | 79 |
| 2000–2007 | 0.341 | 0.296 | 0.364 | 0.888 | 88 |
| 2008–2013 | 0.362 | 0.324 | 0.314 | 0.905 | 66 |
| 2014–2019 | 0.574 | 0.343 | 0.083 | 0.950 | 66 |
| Full period | 0.367 | 0.351 | 0.282 | 0.892 | 299 |
| **Non-EU members** | | | | | |
| 1991–1999 | 0.111 | 0.569 | 0.320 | 0.886 | 70 |
| 2000–2007 | 0.347 | 0.362 | 0.291 | 0.880 | 123 |
| 2008–2013 | 0.340 | 0.374 | 0.286 | 0.910 | 90 |
| 2014–2019 | 0.287 | 0.430 | 0.283 | 0.906 | 90 |
| Full period | 0.286 | 0.420 | 0.293 | 0.894 | 373 |

*Source: Couharde-Grekou (2024) Synthesis Classification. Certainty index = max(P(Fixed), P(Intermediate), P(Floating)). Probabilities may not sum exactly to 1.000 due to rounding.*